\documentclass[%
 aps,
 pre,
 reprint,
 amsmath,
 amssymb,
 superscriptaddress,
 floatfix,
]{revtex4-2}

\usepackage{amsmath}      
\usepackage{amssymb}      
\usepackage{amsfonts}     
\usepackage{mathtools}    
\usepackage{bm}           
\usepackage{physics}      
\usepackage{dsfont}       

\usepackage{graphicx}     
\usepackage{xcolor}       
\usepackage{subcaption}   

\usepackage{booktabs}     
\usepackage{tabularx}     
\usepackage{array}        
\usepackage{multirow}     

\usepackage{hyperref}     
\usepackage[capitalise,noabbrev]{cleveref}  

\usepackage{siunitx}      

\newcolumntype{L}{>{\raggedright\arraybackslash}X}

\hypersetup{
  colorlinks=true,
  linkcolor=blue!50!black,
  citecolor=blue!50!black,
  urlcolor=blue!70!black,
  pdftitle={Non-Markovian Collective Motion from Self-Regulated Perceptual Dynamics},
  pdfauthor={Jyotiranjan Beuria}
}

\begin{document}

\title{Non-Markovian Collective Motion from Self-Regulated
Perceptual Dynamics}

\author{Jyotiranjan Beuria}
\email{jyotiranjan.beuria@gmail.com}
\affiliation{IKS Research Centre, ISS Delhi, Delhi, India}

\date{\today}

\begin{abstract}
Collective motion in active matter is usually modelled through
instantaneous local alignment, where each agent updates its heading from
the current configuration of its neighbours. Many biological and
engineered agents, however, possess internal regulatory variables that
evolve more slowly than alignment itself and can store information about
past alignment states. We introduce a minimal two-timescale model in
which each agent carries a fast perceptual register and a slow regulatory
variable. The fast register encodes the instantaneous tendency to align
with neighbouring headings, while the slow variable integrates recent
alignment and feeds back into subsequent alignment decisions. The internal
dynamics are formulated using a GKSL-derived Bloch representation, used
only as a positivity-preserving effective description of bounded
two-state variables; no microscopic quantum dynamics is assumed. The
model reduces to Vicsek-type alignment in the fast-relaxation,
weak-feedback limit, but shows distinct behaviour when slow feedback is
active. Simulations reveal slow--fast relaxation, feedback-induced
hysteresis, finite memory-dependent loop area, and non-monotonic
coordination between collective order and regulatory tone. These results
show how effective non-Markovian collective motion can emerge from local
internal feedback.
\end{abstract}

\maketitle

\section{Introduction}
\label{sec:introduction}

Collective motion in active matter is a classic example of nonequilibrium
self-organization. From bird flocks
\cite{ballerini2008interaction,bialek2012statistical} and fish schools
\cite{rosenthal2015revealing,couzin2005effective} to self-propelled colloids,
bacterial suspensions, and robotic swarms
\cite{marchetti2013hydrodynamics,bechinger2016active,zhao2018self,rubenstein2014programmable},
local interactions among autonomous units generate ordered phases without
centralized control. The minimal model of this transition, introduced by Vicsek
and collaborators \cite{vicsek1995novel,vicsek2012collective}, treats each agent
as a self-propelled particle that aligns its heading with the average heading of
its neighbours. The resulting order--disorder transition was shown to be
discontinuous by Gr\'{e}goire and Chat\'{e}
\cite{gregorechat2004onset}, and has been characterized through
the Toner--Tu hydrodynamic theory \cite{toner1995long,toner1998flocks} and
many subsequent generalizations \cite{popkin2016physics,sumpter2010collective}.
A common feature of these models is that the alignment rule is effectively
instantaneous: the heading update depends on the present neighbour
configuration, with no internal variable that stores information across
alignment events.

Many natural and engineered agents, however, are not purely reactive. Their
response to a neighbour cue can depend on internal variables that evolve more
slowly than the heading dynamics, such as behavioural persistence, attention,
controller gain, confidence, or adaptive memory. Empirical measurements of
starling flocks show that behavioural inertia measurably delays the propagation
of turning waves through the group \cite{attanasi2014information}, providing
direct evidence that internal timescales shape collective order. Active-matter
models with internal degrees of freedom confirm that such variables can
substantially modify collective behaviour. The inertial spin model of Cavagna
and collaborators \cite{cavagna2015inertial,cavagna2014flocking} introduces an
angular momentum conjugate to the heading and predicts propagating information
waves in flocks. Inertial active Brownian particles \cite{caprini2021inertial}
display collective properties absent in overdamped active particles, and
spontaneous velocity alignment in the overdamped limit arises through a
distinct mechanism \cite{caprini2020spontaneous}. Time-delay variants of the
Vicsek model show that even a fixed actuation lag is sufficient to shift the
collective ordering transition and alter finite-size scaling
\cite{holubec2021finitescaling}, highlighting memory as a generic perturbation
to instantaneous alignment. Memory effects in active glasses and crowded
suspensions \cite{klamser2018thermodynamic,berthier2019glassy} generate
non-Markovian dynamics, while non-Markovian
active baths modify energy equipartition in coupled oscillator systems
\cite{maggi2014generalized}. These examples suggest that slow internal
variables can act as memory channels for fast collective dynamics. Whether
active systems exhibiting such internal degrees of freedom are poised near
criticality \cite{mora2011biological} and what structural features of internal
feedback determine their phase behaviour remain open questions.

Here we develop a minimal active-matter model in which each agent carries two
coupled internal registers. The first is a fast perceptual register that
encodes the agent's instantaneous tendency to align with each neighbour. The
second is a slow regulatory variable, also loosely termed as internal state, that integrates recent perceptual
alignment and feeds back onto the perceptual register, biasing how future
neighbour information is resolved into physical heading. Thus, perception
updates the regulatory state, and the regulatory state modifies subsequent
perception. This closed local loop is the mechanism by which history dependence
enters the collective motion. Decision-making under uncertainty in moving groups
has been modelled using Bayesian and probabilistic internal states
\cite{perezescudero2011collective}; the present framework provides an
operator-algebraic alternative in which internal uncertainty is represented by
transverse Bloch components rather than probability distributions.

The internal dynamics are formulated using an open two-state operator
representation. The longitudinal Bloch component
\(m_z^{ij}\in[-1,1]\) of channel \(j\) at agent \(i\) represents the
resolved align/not-align tendency, while the transverse components
\(m_x^{ij},m_y^{ij}\) represent unresolved competition between the two
alternatives. Relaxation, transverse decay, and amplitude damping are
introduced through a Gorini--Kossakowski--Sudarshan--Lindblad (GKSL)
generator \cite{gorini1976completely,lindblad1976generators}, used here only as
a positivity-preserving effective generator for bounded two-state internal
variables \cite{breuer2002theory}. This is analogous to the role played by
Fokker--Planck or Liouville operators in classical statistical dynamics. The dissipative two-state system with relaxation and
dephasing has a well-established analytical structure \cite{leggett1987dynamics}
that we adapt here to the nonequilibrium active-matter context. No microscopic
quantum dynamics is assumed in the active agents. The operator-state language provides a compact algebraic framework for evolving bounded internal degrees of freedom under switching, relaxation, and
noise.

The model gives four active-matter signatures of internal memory. First, the
usual polar order parameter and the mean local alignment bias grow on the fast
alignment timescale, while the internal regulatory variable relaxes more slowly.
Thus the agents become directionally ordered before their internal memory
variable has fully equilibrated. Second, when the feedback strength is swept up
and then down, the stationary polar order follows different branches. The same
instantaneous feedback value can therefore support different collective states
depending on the previous history of the agents. Third, the area between the
forward and backward branches remains finite over the scanned range of slow
relaxation rates. For the finite sweep protocol used here, this area increases
with the self-relaxation rate, showing that the measured hysteresis
depends not only on the internal memory time but also on the sweep protocol and
relaxation window. We therefore treat the hysteresis area as a
protocol-dependent measure of history dependence, analogous to hysteresis
measurements in driven nonequilibrium systems \cite{cross1993pattern}. Fourth,
the correlation and phase lag between polar order and the internal regulatory
tone are non-monotonic in the feedback strength. This identifies an
intermediate-feedback regime in which the collective motion and the internal
regulatory state are most strongly dynamically locked.

The model is complementary to other internal-state approaches to active matter.
In the inertial spin model \cite{cavagna2014flocking,cavagna2015inertial}, the
slow variable is conjugate to the heading and mediates information propagation.
Here the slow variable is not conjugate to heading; it acts as a regulatory bias
on perceptual alignment and generates memory-dependent hysteresis. The same
mathematical structure can be interpreted differently across systems: in robotic
swarms \cite{rubenstein2014programmable}, the slow variable may represent an
adaptive controller gain or confidence estimate; in biological collectives
\cite{perezescudero2011collective}, it may represent behavioural persistence or
a slow neuromodulatory state; and in earlier operator-state models of
perceptual collective motion \cite{beuria2025collective,beuria2025nonreciprocal},
it corresponds to the slow component of the internal alignment operator. The
framework is therefore substrate-neutral, but quantitative calibration must be
system-specific.

Read as an extension of the Vicsek class \cite{vicsek1995novel,vicsek2012collective},
the present model replaces the instantaneous alignment weight by a dynamical
internal variable generated by perceptual relaxation and modulated by slow
feedback. In the limit of fast transverse decay, fast longitudinal relaxation,
and weak slow feedback, the model reduces to a Vicsek-type local alignment rule
\cite{strombom2011collective}; this reduction is shown in
Appendix~\ref{app:classical_limit}. Away from this limit, the slow regulatory
variable produces an effective memory kernel for the perceptual dynamics
\cite{fodor2016far,maggi2014generalized}, as shown explicitly in
Appendix~\ref{app:memory_kernel}.

The remainder of the paper is organized as follows.
Section~\ref{sec:model} defines the two-register model, derives the mean-field
Bloch equations, and specifies the map from internal state to physical motion.
Section~\ref{sec:observables} introduces the macroscopic observables: polar
order, regulatory tone, self-perception covariance, phase lag, and hysteresis
area. Section~\ref{sec:results} presents the numerical results.
Section~\ref{sec:discussion} discusses interpretation, limitations, and possible
empirical or engineered realizations, and Sec.~\ref{sec:conclusion} concludes.

\section{Model}
\label{sec:model}

We consider $N$ active agents moving in a two-dimensional periodic
domain with a fixed speed $v_0$. Each agent has a physical position
$\mathbf{x}_i(t)$ and an internal state composed of two coupled
registers evolving on separated timescales. The fast register encodes
the agent's instantaneous tendency to align with each of its
perceptually relevant neighbours or environmental cues; the slow
register encodes an internal regulatory variable that integrates the
recent history of the fast register and biases its subsequent
evolution. Throughout this section we use the operator-state language
of driven dissipative two-level systems
\cite{breuer2002theory,leggett1987dynamics} as a positivity-preserving
algebraic framework for the bounded internal variables; the active
agents themselves are classical, and the operator language plays the
same effective role here as Fokker--Planck operators play in classical
stochastic dynamics.

\subsection{Two-register architecture}
\label{sec:architecture}

For each agent $i$, the fast perceptual register consists of channels
$\mathcal{H}_{p,ij}$, one for each perceptually relevant neighbour or
cue $j\in\mathcal{N}_i$. Each channel is a two-state register with
basis states $\ket{1}_{p,ij}$ and $\ket{0}_{p,ij}$ representing,
respectively, the agent's instantaneous tendency to align with that
neighbour or cue and to remain unaligned. The slow regulatory register
$\mathcal{H}_{s,i}$ is also represented as a two-state register, with
basis states $\ket{1}_{s,i}$ and $\ket{0}_{s,i}$ representing high and
low internal regulatory engagement. Depending on the system, the slow
register's state can correspond to an adaptive controller gain, a
confidence variable, a logged memory state, a slow neuromodulatory
level, or any other internal degree of freedom that accumulates
information over a longer timescale than the fast alignment dynamics.

The composite Hilbert space for agent $i$ is
\begin{equation}
\mathcal{H}_{i}
=
\mathcal{H}_{s,i}
\otimes
\bigotimes_{j\in\mathcal{N}_i}
\mathcal{H}_{p,ij},
\qquad
\mathcal{H}_{s,i},\mathcal{H}_{p,ij}\simeq\mathbb{C}^{2}.
\label{eq:hilbert}
\end{equation}
The tensor-product structure represents the internal factorization of
\emph{a single agent's} state into its slow and fast registers; it does
not represent multiple agents. Inter-agent coupling enters separately
through an effective neighbour field derived in
Sec.~\ref{sec:bloch_reduction} below. The internal state of each agent
is described by a reduced density operator $\rho_i$ acting on
$\mathcal{H}_i$ and is propagated under an open-system generator
introduced next.

\subsection{Open-system formulation}
\label{sec:gksl}

The internal state of each agent is not treated as an isolated dynamical
system. At every time, the agent receives information from neighbouring
agents, external cues, and unresolved environmental or computational
degrees of freedom. We therefore describe the internal registers of each
agent as an effective open system. The purpose of this formulation is not
to assume microscopic quantum dynamics, but to obtain a compact
positivity-preserving evolution equation for bounded two-state internal
variables.

Let \(\rho_i(t)\) denote the reduced internal state of agent \(i\). The
diagonal entries of \(\rho_i\) represent occupation probabilities of
the internal alternatives, such as align/not-align in the fast register
or low/high regulatory engagement in the slow register. The off-diagonal
entries represent unresolved competition between the two alternatives.
These off-diagonal components are not interpreted as microscopic quantum
coherences of the agent; they are auxiliary dynamical variables that
allow the two-state internal system to relax, switch, and lose unresolved
competition in a controlled way.

We evolve \(\rho_i(t)\) using a Gorini--Kossakowski--Sudarshan--Lindblad
(GKSL) generator \cite{gorini1976completely,lindblad1976generators},
\begin{equation}
\dot{\rho}_i
=
-i\big[H_i,\rho_i\big]
+
\sum_{\alpha}
\left(
L_{\alpha,i}\rho_i L_{\alpha,i}^{\dagger}
-
\frac{1}{2}
\big\{
L_{\alpha,i}^{\dagger}L_{\alpha,i},\rho_i
\big\}
\right).
\label{eq:gksl}
\end{equation}
The first term, \(-i[H_i,\rho_i]\), gives deterministic evolution in the
internal model space. In the Bloch representation used below, it produces
precession-like switching between internal alternatives. The second term
contains the Lindblad operators \(L_{\alpha,i}\), which encode effective
noise, relaxation, and decay of unresolved internal competition. This
form guarantees that \(\rho_i(t)\) remains trace-preserving and positive
during the evolution.

\subsubsection{Hamiltonian terms}
\label{sec:Hamiltonian}

The internal Hamiltonian of agent \(i\) is decomposed into four parts,
\begin{equation}
H_i
=
H_s+H_p+H_{sp}+P\big(X(t)\big),
\label{eq:H_total}
\end{equation}
where \(H_s\) acts on the slow regulatory register, \(H_p\) acts on the
fast perceptual registers, \(H_{sp}\) couples the slow and fast
registers, and \(P(X)\) represents an optional external drive. Explicitly,
\begin{align}
H_s
&=
-\Gamma_s S_x
-
J_s S_z,
\label{eq:Hs}
\\[2pt]
H_p
&=
-\Gamma
\sum_{j\in\mathcal{N}_i}
\sigma_x^{(ij)}
-
\sum_{\substack{j,k\in\mathcal{N}_i\\ j<k}}
J_{jk}
\sigma_z^{(ij)}
\sigma_z^{(ik)},
\label{eq:Hp}
\\[2pt]
H_{sp}
&=
-
\sum_{j\in\mathcal{N}_i}
\kappa_j
S_z\otimes\sigma_z^{(ij)},
\label{eq:Hsp}
\\[2pt]
P(X)
&=
-
\sum_{j\in\mathcal{N}_i}
h_{ij}\big(X(t)\big)
\sigma_z^{(ij)}.
\label{eq:Pdrive}
\end{align}
Here \(S_\alpha\) are Pauli operators acting on the slow register of
agent \(i\), while \(\sigma_\alpha^{(ij)}\) are Pauli operators acting on
the fast perceptual channel associated with neighbour or cue \(j\). $\mathcal{N}_i$ is the set of neighbours for agent $i$.

The slow-register Hamiltonian \(H_s\) contains two terms. The transverse
term \(-\Gamma_s S_x\) allows the slow register to switch between its two
states, which we interpret as low and high regulatory engagement. The
parameter \(\Gamma_s\) therefore sets an internal exploration rate for
the slow register. The longitudinal term \(-J_s S_z\) is a baseline bias:
if \(J_s>0\), it energetically favours one regulatory state over the
other. In the simulations below, the main memory timescale of the slow
register is controlled by the dissipative relaxation rate
\(\gamma_s\), introduced in Sec.~\ref{sec:dissipative_channels}. The
intended slow--fast regime is that \(\gamma_s^{-1}\) should be much longer than the
fast perceptual relaxation times, so that the slow register stores a
history of fast alignment over many perceptual updates.

The fast-register Hamiltonian \(H_p\) describes the internal alignment
decision of agent \(i\) with respect to its neighbours. The transverse
term
\[
-\Gamma\sum_{j\in\mathcal{N}_i}\sigma_x^{(ij)}
\]
drives switching between the two alternatives of each perceptual
channel: align with neighbour \(j\), or do not align with it. Thus
\(\Gamma\) sets the fast switching rate. The second term,
\[
-
\sum_{\substack{j,k\in\mathcal{N}_i\\ j<k}}
J_{jk}\sigma_z^{(ij)}\sigma_z^{(ik)},
\]
couples different perceptual channels of the same agent. Its role is to
make the agent's alignment tendencies internally consistent. For positive \(J_{jk}\), compatible perceptual channels reinforce each
other: a bias toward aligning with one cue increases the tendency to
align with another cue that supports the same heading choice.
In the mean-field reduction below this channel--channel coupling becomes
an average local alignment field.

The slow--fast coupling \(H_{sp}\) is the key internal feedback term. It
couples the longitudinal state of the slow register, \(S_z\), to the
longitudinal align/not-align variable \(\sigma_z^{(ij)}\) of each fast
channel. The coupling strength \(\kappa_j\) specifies how strongly the slow
regulatory state influences the fast perceptual channel associated with
cue \(j\). For \(\kappa_j>0\), a positive slow-register value,
\(\langle S_z\rangle>0\), increases the tendency of that channel to
resolve cue \(j\) as an align decision. Thus the slow register acts like
an internal gain on perceptual alignment: when the regulatory state is
high, neighbour cues are more easily converted into alignment weights.
This is the slow-to-fast part of the internal feedback loop. The reverse
fast-to-slow part is introduced phenomenologically in
Eq.~\eqref{eq:s_dynamics}, where the mean fast-register alignment updates
the slow regulatory variable.

Finally, the drive term \(P(X)\) represents an external control or
stimulus. The scalar field \(h_{ij}(X(t))\) biases the fast channel
\(ij\) toward or away from alignment depending on the external variable
\(X(t)\). This term may represent, for example, a target, environmental
cue, or externally controlled perturbation. In the numerical simulations
reported in Sec.~\ref{sec:results}, this external drive is set to zero
unless an explicit response to external forcing is being probed.

\subsubsection{Dissipative channels}
\label{sec:dissipative_channels}

The Hamiltonian terms describe deterministic internal switching and
biases. To model relaxation and loss of unresolved internal competition,
we add three dissipative channels through the Lindblad part of
Eq.~\eqref{eq:gksl}.

\paragraph*{Fast-register dephasing.}
For each fast channel \(ij\), we include
\begin{equation}
L^{(\phi)}_{ij}
=
\sqrt{\gamma_\phi}\;
I_s\otimes\sigma_z^{(ij)}.
\label{eq:Lphi}
\end{equation}
This operator damps the off-diagonal elements of the fast-register
density matrix while leaving the occupation probabilities unchanged. In
physical terms, it removes unresolved competition between the align and
not-align alternatives without directly favouring either one. It
therefore contributes to the transverse decay time \(T_2\) introduced
in the Bloch equations.

\paragraph*{Fast-register amplitude damping.}
Population relaxation between the two fast-register alternatives is
modelled by
\begin{equation}
L^-_{ij}
=
\sqrt{\gamma_-}\;
I_s\otimes\sigma_-^{(ij)},
\qquad
L^+_{ij}
=
\sqrt{\gamma_+}\;
I_s\otimes\sigma_+^{(ij)} .
\label{eq:Lpm}
\end{equation}
The operator \(L^-_{ij}\) transfers population from the align state to
the not-align state, while \(L^+_{ij}\) transfers population in the
opposite direction. Together they relax the longitudinal component
\(m_z^{ij}\) toward a field-dependent equilibrium value. These rates
define the longitudinal relaxation time \(T_1\), which controls how fast
the resolved align/not-align bias adapts to the effective field.

\paragraph*{Slow-register relaxation.}
The slow regulatory register relaxes through
\begin{equation}
L^-_{s,i}
=
\sqrt{\gamma_s}\;
S_-\otimes I_P .
\label{eq:Ls}
\end{equation}
This channel drives the slow register back toward its baseline state at
rate \(\gamma_s\). Hence \(\gamma_s^{-1}\) is the memory horizon of the
slow variable: small \(\gamma_s\) means slow forgetting and long memory,
whereas large \(\gamma_s\) means rapid relaxation and short memory. This
timescale is the main slow--fast control parameter in the model.

Together, these dissipative channels implement the minimal internal
physics required for the model. The fast register can switch between
alignment alternatives, lose unresolved competition, and relax toward a
field-dependent resolved choice. The slow register can store the recent
history of the fast register and relax back to baseline. The combination
of the slow-to-fast bias in \(H_{sp}\) and the fast-to-slow feedback in
Eq.~\eqref{eq:s_dynamics} produces the closed internal feedback loop
responsible for the memory-dependent collective dynamics studied below.

\subsection{Bloch reduction and mean-field equations}
\label{sec:bloch_reduction}

Direct integration of Eq.~\eqref{eq:gksl} on the full Hilbert space
of Eq.~\eqref{eq:hilbert} requires evolving a density matrix of
dimension $2^{|\mathcal{N}_i|+1}\times 2^{|\mathcal{N}_i|+1}$ for
each agent, which is prohibitive for the ensembles of hundreds to
thousands of agents we wish to simulate. We therefore project the
GKSL dynamics onto the Pauli basis and adopt a mean-field
factorization, obtaining coupled ordinary differential equations for
the Bloch components of each register.

For the fast channel $ij$, define
\begin{equation}
m_\alpha^{ij}(t)
=
\langle \sigma_\alpha^{(ij)}\rangle
=
\mathrm{Tr}\!\left[
\rho_i\,\sigma_\alpha^{(ij)}
\right],
\quad
\alpha\in\{x,y,z\}.
\label{eq:m_def}
\end{equation}
The longitudinal component $m_z^{ij}\in[-1,1]$ encodes the resolved
alignment bias toward cue $j$, while the transverse components
$m_x^{ij},m_y^{ij}$ describe unresolved competition between the
align and not-align alternatives; the latter relax rapidly under the
dephasing channel of Eq.~\eqref{eq:Lphi} and play no direct role in
the heading map. For the slow register of agent $i$, define
\begin{equation}
s_\alpha^i(t)
=
\langle S_\alpha^{(i)}\rangle
=
\mathrm{Tr}\!\left[
\rho_i\, S_\alpha^{(i)}
\right].
\label{eq:s_def}
\end{equation}
We use the short notation $s_i\equiv s_z^i$ for the slow longitudinal
component, which is the only slow-register variable that enters the
fast dynamics through the coupling in Eq.~\eqref{eq:Hsp}.

\paragraph*{Mean-field factorization.}
We approximate the reduced state of agent $i$ as a product of single-
register reduced states,
\begin{equation}
\rho_i(t)
\approx
\rho_{s,i}(t)
\otimes
\bigotimes_{j\in\mathcal{N}_i}
\rho_{p,ij}(t),
\label{eq:mf_ansatz}
\end{equation}
with each factor in standard Bloch form,
\begin{equation}
\rho_{p,ij}=\tfrac{1}{2}\!\left(I+\mathbf{m}^{ij}\!\cdot\boldsymbol{\sigma}^{(ij)}\right),
\;
\rho_{s,i}=\tfrac{1}{2}\!\left(I+\mathbf{s}^{i}\!\cdot\mathbf{S}^{(i)}\right).
\label{eq:Bloch_form}
\end{equation}
This factorization neglects connected correlations between fast
channels of the same agent and between the fast and slow registers.
It is expected to be accurate when each agent integrates over several
neighbours and when the fast couplings $J_{jk}$ and $\kappa_j$ are
moderate; the limitations of this closure are discussed in
Sec.~\ref{sec:discussion}. Component equations follow from
$\dot{m}_\alpha^{ij}=\mathrm{Tr}[\sigma_\alpha^{(ij)}\dot{\rho}_i]$
and $\dot{s}_\alpha^i=\mathrm{Tr}[S_\alpha^{(i)}\dot{\rho}_i]$.

\paragraph*{Effective field on a fast channel.}
Under the mean-field closure, the fast channel $ij$ experiences a
longitudinal effective field
\begin{equation}
h_{ij}^{\mathrm{eff}}(t)
=
h_{ij}\!\big(X(t)\big)
+
J\,\bar{m}_z^{\,\mathcal{N}_i}(t)
+
\kappa\, s_i(t),
\label{eq:heff_methods}
\end{equation}
\begin{equation}
\bar{m}_z^{\,\mathcal{N}_i}(t)
=
\frac{1}{|\mathcal{N}_i|}
\sum_{j\in\mathcal{N}_i} m_z^{ij}(t),
\label{eq:mzbar}
\end{equation}
where the first term is the external drive, the second couples the
fast channel to the average longitudinal alignment of the agent's
neighbours through the coupling strength $J$, and the third is the
slow-register bias with strength $\kappa$. Equation~\eqref{eq:heff_methods}
collapses the Ising-like consistency term of Eq.~\eqref{eq:Hp} into a
neighbour-averaged field, which is the form actually used in the
large-ensemble simulations of Sec.~\ref{sec:results}; we use a
single coupling constant $\kappa$ rather than channel-specific
$\kappa_j$, since the simulations do not require differentiation
between cues.

\paragraph*{Unitary precession.}
The commutator part of Eq.~\eqref{eq:gksl}, evaluated on the single channel Hamiltonian
$H_{\mathrm{eff}}^{(ij)}=-\Gamma\sigma_x^{(ij)}-h_{ij}^{\mathrm{eff}}\sigma_z^{(ij)}$,
yields Bloch precession around the effective field
$\mathbf{B}_{ij}=(-\Gamma,0,-h_{ij}^{\mathrm{eff}})$,
\begin{equation}
\dot{\mathbf{m}}^{ij}\big|_{\mathrm{unitary}}
=
2\,\mathbf{B}_{ij}\times\mathbf{m}^{ij}.
\label{eq:precession}
\end{equation}

\paragraph*{Dissipative relaxation.}
The Lindblad channels of Sec.~\ref{sec:dissipative_channels} produce
the standard relaxation structure
\begin{equation}
\frac{1}{T_1}
=
\gamma_-+\gamma_+,
\qquad
\frac{1}{T_2}
=
2\gamma_\phi+\frac{1}{2T_1},
\label{eq:T1T2}
\end{equation}
where $T_1$ is the longitudinal relaxation time and $T_2$ the
transverse decay time. The amplitude-damping channels drive the
longitudinal component toward a field-dependent equilibrium, written
phenomenologically as
\begin{equation}
m_z^{\mathrm{eq}}(h)
=
\tanh(\beta h),
\label{eq:mzeq}
\end{equation}
where $\beta$ is a noise-sharpness parameter set by the ratio
$\gamma_-/\gamma_+$ and the bath temperature in the open-system
microscopic derivation. For our purposes $\beta$ is a free parameter
controlling how sharply the resolved alignment tracks the local
field.

\paragraph*{Fast-register Bloch equations.}
Combining the unitary and dissipative contributions yields the
mean-field Bloch equations for each fast channel,
\begin{subequations}
\label{eq:bloch}
\begin{align}
\dot{m}_x^{ij}
&=
2 h_{ij}^{\mathrm{eff}}\, m_y^{ij}
-
\frac{m_x^{ij}}{T_2},
\label{eq:bloch_x}\\
\dot{m}_y^{ij}
&=
-2 h_{ij}^{\mathrm{eff}}\, m_x^{ij}
+
2\Gamma\, m_z^{ij}
-
\frac{m_y^{ij}}{T_2},
\label{eq:bloch_y}\\
\dot{m}_z^{ij}
&=
-2\Gamma\, m_y^{ij}
-
\frac{m_z^{ij}-\tanh(\beta\,h_{ij}^{\mathrm{eff}})}{T_1}.
\label{eq:bloch_z}
\end{align}
\end{subequations}
The detailed derivation of these equations from the GKSL generator
is given in Appendix~\ref{app:bloch_derivation}.

\subsection{Slow regulatory dynamics}
\label{sec:slow_dynamics}

The slow regulatory variable of agent \(i\) is the longitudinal
component
\[
s_i(t)\equiv s_z^i(t),
\qquad
s_i\in[-1,1].
\]
It represents an internal state that changes more slowly than the
fast perceptual register and modulates how future neighbour cues are
resolved. Depending on the system, \(s_i\) may correspond to an
adaptive controller gain, a confidence variable, a local memory state,
or a slowly varying behavioural response.

We model this variable by
\begin{equation}
\dot{s}_i
=
-\gamma_s\,(s_i-s_{\mathrm{eq}})
+
\lambda_{\rm fb}\,(1-s_i^2)\,
\bar{m}_z^{\,\mathcal{N}_i}(t),
\label{eq:s_dynamics}
\end{equation}
where
\begin{equation}
\bar{m}_z^{\,\mathcal{N}_i}(t)
=
\frac{1}{|\mathcal{N}_i|}
\sum_{j\in\mathcal{N}_i}
m_z^{ij}(t)
\end{equation}
is the mean fast-register alignment bias of agent \(i\). Positive
\(\bar{m}_z^{\,\mathcal{N}_i}\) means that the agent's fast perceptual
channels are, on average, biased toward alignment with its neighbours.

The first term in Eq.~\eqref{eq:s_dynamics},
\(-\gamma_s(s_i-s_{\mathrm{eq}})\), relaxes the slow variable back to
its baseline value \(s_{\mathrm{eq}}\). Thus \(\gamma_s\) is the
self-relaxation or forgetting rate of the slow register, and
\(\gamma_s^{-1}\) is its characteristic memory time. Smaller
\(\gamma_s\) corresponds to slower forgetting, while larger
\(\gamma_s\) corresponds to faster return to baseline.

The second term is the feedback from the fast perceptual register to the
slow regulatory variable. The coefficient \(\lambda_{\rm fb}\) controls
how strongly recent perceptual alignment updates \(s_i\). For positive
\(\lambda_{\rm fb}\), sustained positive alignment bias increases
\(s_i\), which in turn biases later perceptual resolution through the
slow-to-fast coupling in Eq.~\eqref{eq:Hsp}. The factor
\((1-s_i^2)\) prevents unbounded growth: as \(s_i\) approaches either
endpoint of the interval \([-1,1]\), the feedback term weakens and the
slow variable saturates.

The intended regime is
\begin{equation}
\gamma_s^{-1}\gg T_1,\,T_2,\,\Gamma^{-1},
\label{eq:hierarchy}
\end{equation}
so that the slow register evolves over many fast perceptual relaxation
cycles. If the state of the system is specified by the full set of
variables
\[
\Big\{
\mathbf{x}_i,\hat{\mathbf p}_i,s_i,
m_x^{ij},m_y^{ij},m_z^{ij}
\Big\}_{i=1,\ldots,N;\,j\in\mathcal N_i},
\]
then the dynamics are Markovian: the time derivative of each variable is
determined by the current values of these variables. No additional past
trajectory is required.

However, if the slow variable \(s_i(t)\) is not kept as an explicit
state variable, its effect on the fast register must be represented
through the past history of the fast-register alignment. In that reduced
description, the present effective field acting on \(m_z^{ij}(t)\)
depends not only on the current fast variables but also on their earlier
values, because \(s_i(t)\) has accumulated them over its relaxation time
\(\gamma_s^{-1}\). Thus the full system is Markovian in the enlarged
state space, while the fast perceptual dynamics alone are effectively
non-Markovian. The explicit memory-kernel form of this reduced dynamics
is derived in Appendix~\ref{app:memory_kernel}.

\subsection{From internal state to physical motion}
\label{sec:motion}

The internal Bloch variables drive physical motion through a heading
map that converts the longitudinal alignment bias of each perceptual
channel into a weight on the heading direction of the corresponding
neighbour. Let
\(\hat{\mathbf p}_i(t)\) denote the unit heading direction of agent
\(i\), and let \(\hat{\mathbf p}_j(t)\) denote the heading direction
of a neighbour \(j\in\mathcal{N}_i\). In the simulations below we use
a topological neighbourhood rule: each agent interacts with its
\(n\) nearest neighbours and all neighbour cues are assigned equal
weight,
\begin{equation}
w_{ij}
=
\frac{1}{n},
\qquad
j\in\mathcal{N}_i .
\end{equation}
Agents outside \(\mathcal{N}_i\) have zero weight.

The unnormalised desired heading of agent \(i\) is defined as
\begin{equation}
\mathbf{u}_i(t)
=
\rho\,\hat{\mathbf p}_i(t)
+
\sum_{j\in\mathcal{N}_i}
w_{ij}\,
\frac{1+m_z^{ij}(t)}{2}\,
\hat{\mathbf p}_j(t).
\label{eq:desired_heading}
\end{equation}
Here \(\rho\geq0\) is a persistence parameter that weights the
agent's current heading. The factor
\((1+m_z^{ij})/2\in[0,1]\) converts the longitudinal Bloch component
of the perceptual channel \(ij\) into an alignment weight:
\(m_z^{ij}=+1\) gives full weight to the heading of neighbour \(j\),
\(m_z^{ij}=-1\) suppresses that neighbour's contribution, and
\(m_z^{ij}=0\) assigns intermediate weight.

The desired heading direction is
\begin{equation}
\hat{\mathbf p}^{\,\mathrm{des}}_i(t)
=
\frac{\mathbf{u}_i(t)}
{|\mathbf{u}_i(t)|+\epsilon},
\label{eq:desired_unit_heading}
\end{equation}
where \(\epsilon>0\) is a numerical regulariser preventing division by
zero when \(|\mathbf u_i|\) is transiently small. The physical heading
relaxes toward this desired heading according to
\begin{equation}
\dot{\hat{\mathbf p}}_i(t)
=
\frac{
\hat{\mathbf p}^{\,\mathrm{des}}_i(t)
-
\hat{\mathbf p}_i(t)
}{\tau_p},
\label{eq:heading_relaxation}
\end{equation}
with \(\tau_p\) the heading-relaxation time. The physical position then
evolves as
\begin{equation}
\dot{\mathbf{x}}_i(t)
=
v_0\,\hat{\mathbf p}_i(t),
\label{eq:velocity}
\end{equation}
where \(v_0\) is the constant agent speed. After each numerical update,
\(\hat{\mathbf p}_i\) is renormalised to unit length. The regulariser
\(\epsilon\) is held fixed across all simulations at a value much
smaller than the typical \(|\mathbf u_i|\); varying it within this
small range does not affect the reported macroscopic trends.

This heading map makes the model a direct extension of Vicsek-type
polar alignment. In the limit of fast transverse decay, fast
longitudinal relaxation, and vanishing slow feedback, the fast Bloch
equations \eqref{eq:bloch} adiabatically eliminate the transverse
components and drive
\begin{equation}
m_z^{ij}
\to
\tanh\!\left(
\beta h_{ij}^{\mathrm{eff}}
\right).
\end{equation}
In the high-sharpness limit \(\beta\to\infty\), this becomes a binary
neighbour-following weight. Equation~\eqref{eq:desired_heading} then
reduces to a weighted Vicsek-like heading update, with the alignment
weights generated dynamically by the internal perceptual variables. The
slow-register feedback term in Eq.~\eqref{eq:s_dynamics} is the
ingredient that introduces history dependence beyond an instantaneous
alignment rule: it contributes to \(h_{ij}^{\mathrm{eff}}\) through the
\(\kappa s_i\) term, with a memory horizon set by \(\gamma_s^{-1}\).

The model is fully specified by Eqs.~\eqref{eq:bloch},
\eqref{eq:s_dynamics}, \eqref{eq:desired_heading},
\eqref{eq:heading_relaxation}, and \eqref{eq:velocity}, together with
the prescription for the neighbourhood \(\mathcal{N}_i\). The
parameters of the model and their operational meanings are summarized
in Table~\ref{tab:parameters}.

\begin{table}[!t]
\caption{Model parameters and their operational meaning. The slow--fast
hierarchy required for the results of Sec.~\ref{sec:results} is
\(\gamma_s^{-1}\gg T_1,T_2,\Gamma^{-1}\).}
\label{tab:parameters}
\centering
\begin{ruledtabular}
\begin{tabular}{ll}
Symbol & Operational meaning \\
\hline
\(\Gamma\)              & Fast-register switching rate \\
\(T_1\)                 & Longitudinal perceptual relaxation time \\
\(T_2\)                 & Transverse decay time \\
\(J\)                   & Heading-evidence coupling in \(h_{ij}^{\mathrm{eff}}\) \\
\(\kappa\)              & Slow-to-fast bias strength \\
\(\lambda\)             & Fast-to-slow feedback strength \\
\(\gamma_s\)            & Slow-register relaxation rate \\
\(s_{\mathrm{eq}}\)     & Slow-register baseline \\
\(\beta\)               & Inverse noise or choice sharpness \\
\(\rho\)                & Heading-persistence weight \\
\(\tau_p\)              & Heading-relaxation time \\
\(v_0\)                 & Agent speed \\
\(\epsilon\)            & Heading regulariser \\
\(w_{ij}\)              & Sensory weight on neighbour \(j\) \\
\end{tabular}
\end{ruledtabular}
\end{table}

\section{Observables}
\label{sec:observables}

We characterize the collective behaviour of the model through four
macroscopic observables, all computed from the trajectories defined in
Sec.~\ref{sec:model}: the polar order parameter \(M(t)\), the
ensemble-averaged slow-register tone \(S(t)\), the temporal covariance
and phase lag \((C_{MS},\Phi_{MS})\) between these two quantities, and
the hysteresis area \(\mathcal{A}_{\mathrm{hyst}}\) under cyclic
feedback sweeps.

\subsection{Collective alignment and slow-register tone}
\label{sec:order_params}

The macroscopic alignment of the ensemble is measured by the standard
polar order parameter
\begin{equation}
M(t)
=
\left|
\frac{1}{N}
\sum_{i=1}^{N}
\hat{\mathbf p}_i(t)
\right|.
\label{eq:order_param}
\end{equation}
Equivalently, writing
\(\hat{\mathbf p}_i(t)=(p_{ix}(t),p_{iy}(t))\), this is
\begin{equation}
M(t)
=
\sqrt{
\left(
\frac{1}{N}\sum_{i=1}^{N}p_{ix}(t)
\right)^2
+
\left(
\frac{1}{N}\sum_{i=1}^{N}p_{iy}(t)
\right)^2
}.
\end{equation}
The values \(M(t)\to1\) and \(M(t)\to0\) correspond, respectively, to
global polar alignment and disordered motion. Equation
\eqref{eq:order_param} is the usual Vicsek-type polar order parameter,
but here the heading vectors \(\hat{\mathbf p}_i(t)\) are generated by
the internal Bloch dynamics and the slow regulatory feedback through
Eqs.~\eqref{eq:desired_heading}--\eqref{eq:heading_relaxation}.

The internal regulatory state of the ensemble is summarized by the
ensemble-averaged slow-register tone
\begin{equation}
S(t)
=
\frac{1}{N}
\sum_{i=1}^{N} s_i(t),
\qquad
S(t)\in[-1,1],
\label{eq:S_def}
\end{equation}
where \(s_i(t)\) is the slow longitudinal component of agent \(i\)
introduced in Sec.~\ref{sec:slow_dynamics}. Positive values
\(S(t)>0\) correspond to a population in which the slow register is
biased toward the high-engagement or high-gain branch and therefore
biases the fast-register field toward alignment. Values
\(S(t)\approx0\) correspond to a neutral baseline, while negative values
indicate a low-gain or externally reactive regime. The pair
\((M(t),S(t))\) provides the two-dimensional macroscopic state space in
which the feedback-controlled ordering and hysteresis reported in
Sec.~\ref{sec:results} are visualised.

\subsection{Self--perception coupling and phase lag}
\label{sec:correlation_obs}

The dynamical coupling between the fast and slow registers at the
collective level is quantified by the normalized temporal covariance
of $M(t)$ and $S(t)$,
\begin{equation}
C_{MS}
=
\frac{
\big\langle\,
\big[M(t)-\bar{M}\big]\big[S(t)-\bar{S}\big]
\,\big\rangle_t
}{
\sigma_M\,\sigma_S
},
\qquad
C_{MS}\in[-1,1],
\label{eq:CMS}
\end{equation}
where $\bar{M},\bar{S}$ and $\sigma_M,\sigma_S$ are the time-averaged
means and standard deviations evaluated over the post-transient
window. Positive $C_{MS}$ indicates that fluctuations of $M(t)$ and
$S(t)$ are positively entrained; vanishing $C_{MS}$ indicates
decoupled fluctuations.

A complementary measure is the average phase lag $\Phi_{MS}$ extracted
from the analytic signals of $M(t)$ and $S(t)$. Defining
$\tilde{X}(t)=X(t)+i\,\mathcal{H}[X(t)]$, where $\mathcal{H}[\cdot]$
denotes the Hilbert transform, the instantaneous phase difference is
\begin{equation}
\Delta\Phi_{MS}(t)
=
\arg\big[\tilde{M}(t)\big]
-
\arg\big[\tilde{S}(t)\big],
\label{eq:phase_diff}
\end{equation}
and $\Phi_{MS}$ is its time average over the post-transient window.
The analytic-signal phase is well defined when $M(t)$ and $S(t)$
carry oscillatory or relaxational content, which is the case during
the cyclic-feedback protocol of
Sec.~\ref{sec:hysteresis_results}.  A
positive $\Phi_{MS}$ indicates that fluctuations in $S(t)$ lag those
in $M(t)$, which is the expected behaviour for a slow integrative
variable that responds to fast-register alignment.

\subsection{Hysteresis area}
\label{sec:hyst_obs}

We quantify memory under cyclic forcing by measuring hysteresis in the
stationary order parameter during slow sweeps of the feedback strength.
The sweep protocol is
\begin{equation}
\lambda_{\rm fb}(t):
\;
0
\;\longrightarrow\;
\lambda_{{\rm fb};\max}
\;\longrightarrow\;
0,
\label{eq:lambda_sweep}
\end{equation}
where the sweep is slow enough that the system can approach a
quasi-stationary state at each value of \(\lambda_{\rm fb}\). Along the
increasing branch and the decreasing branch we separately record the
post-transient averages
\[
M^\ast(\lambda_{\rm fb})
=
\langle M(t)\rangle_{\rm post},
\qquad
S^\ast(\lambda_{\rm fb})
=
\langle S(t)\rangle_{\rm post}.
\]
If the system has no memory, the two branches coincide. If the slow
regulatory state carries history, the same value of
\(\lambda_{\rm fb}\) can give different stationary orders on the
forward and backward sweeps.

The hysteresis area is defined as the absolute separation between the
two branches:
\begin{equation}
\mathcal{A}_{\mathrm{hyst}}
=
\int_{0}^{\lambda_{{\rm fb};\max}}
\left|
M^\ast_{\mathrm{fwd}}(\lambda_{\rm fb})
-
M^\ast_{\mathrm{bwd}}(\lambda_{\rm fb})
\right|
\,d\lambda_{\rm fb}.
\label{eq:hyst_area}
\end{equation}
This definition is independent of the direction in which the loop is
traversed and vanishes when the forward and backward branches are the
same. The corresponding loop in the \((S^\ast,M^\ast)\) plane can also
be plotted, but its oriented area depends on the sense of traversal. We
therefore use Eq.~\eqref{eq:hyst_area} as the primary scalar measure of
history dependence.

\begin{figure*}[!ht]
    \centering
    \includegraphics[width=0.48\linewidth]{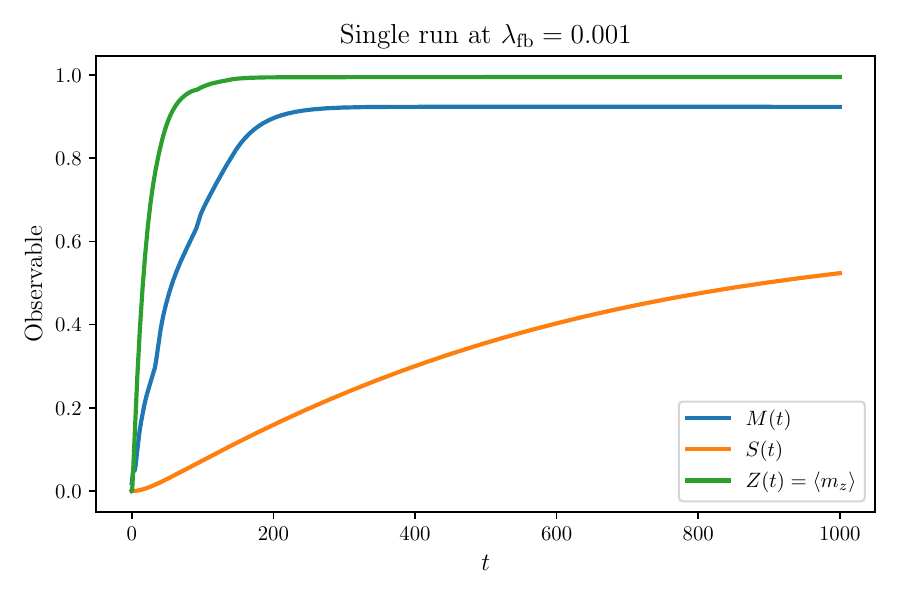}
    \includegraphics[width=0.48\linewidth]{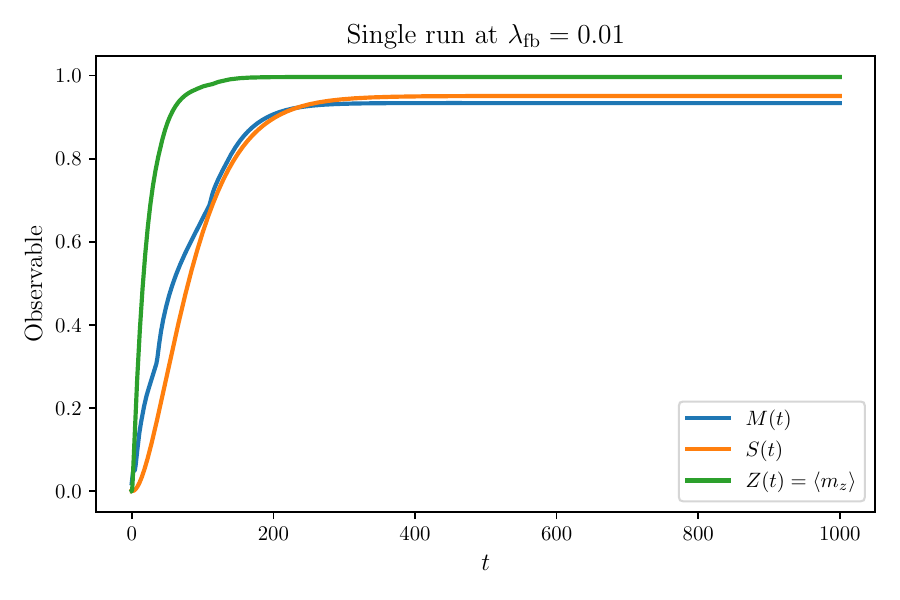}
    \caption{Representative slow--fast relaxation dynamics. The polar
    order parameter \(M(t)\), slow-register tone \(S(t)\), and mean
    perceptual bias \(Z(t)=\langle m_z\rangle\) are shown for
    \(\lambda_{\rm fb}=10^{-3}\) (left) and
    \(\lambda_{\rm fb}=10^{-2}\) (right). The perceptual bias and polar
    order rise rapidly, whereas the regulatory variable evolves on a
    slower timescale. Increasing \(\lambda_{\rm fb}\) accelerates the
    growth of \(S(t)\), demonstrating direct feedback from perceptual
    alignment to the slow register.}
    \label{fig:single_run_relaxation}
\end{figure*}

The dimensionless quantity
\(\mathcal{A}_{\mathrm{hyst}}\) measures how strongly the collective
order depends on the past regulatory state. It is not a universal
constant: it depends on the slow relaxation rate \(\gamma_s\), the sweep
range, the relaxation time allowed at each sweep point, and the feedback
couplings. In this work we use it as a protocol-dependent memory
observable. A nonzero value indicates that the slow internal variable
stores enough history to make the reduced collective dynamics
effectively non-Markovian.
\section{Numerical results}
\label{sec:results}

We now analyse the collective dynamics generated by the coupled
fast-register and slow-register equations. The purpose of the numerical
study is to isolate the physical consequences of the slow internal
feedback loop: separation of timescales, feedback-controlled ordering,
hysteresis under cyclic feedback, and self--perception entrainment. All
simulations use a fixed topological neighbourhood graph with uniform
weights \(w_{ij}=1/n\), and the polar order parameter is computed from
the physical headings according to Eq.~\eqref{eq:order_param}. The
equations are integrated with a fourth-order Runge--Kutta scheme with
time step \(\Delta t=0.05\).

In all simulations, we have taken uniform random distribution of headings of the agents as the initial configurations.
Unless otherwise stated, the baseline parameters are
\(\Gamma=0.1\), \(T_1=20\), \(T_2=8\), \(\kappa=1\),
\(\beta=6\), \(\gamma_s=10^{-3}\), \(\rho=0.7\),
\(\tau_p=0.25\), with \(s_{\rm eq}=h_{\rm ext}=0\) and \(v_0=1\).
Figure~\ref{fig:single_run_relaxation} uses \(N=400\), \(n=6\), and
\(J=3\), while Figs.~\ref{fig:hysteresis_lambda}--
\ref{fig:self_perception_coupling} use \(n=4\).
The smaller neighbourhood in the sweep simulations is chosen only to
make the branch separation easier to visualize; the underlying
slow--fast mechanism is the same.

\subsection{Slow--fast relaxation and emergence of collective order}
\label{sec:relaxation_results}

Figure~\ref{fig:single_run_relaxation} shows representative time
evolution of the polar order parameter \(M(t)\), the slow-register tone
\(S(t)\), and the mean perceptual bias
\[
Z(t)
=
\frac{1}{N}
\sum_{i=1}^{N}
\frac{1}{|\mathcal N_i|}
\sum_{j\in\mathcal N_i}
m_z^{ij}(t).
\]
The two feedback values shown, \(\lambda_{\rm fb}=10^{-3}\) and
\(\lambda_{\rm fb}=10^{-2}\), illustrate the separation between the fast
perceptual layer and the slower regulatory layer.

For both feedback values, the perceptual variable \(Z(t)\) rises rapidly
toward its high-alignment branch. The polar order parameter \(M(t)\) also
increases from its initially disordered value to a high ordered state,
showing that the internal alignment weights are successfully converted
into physical polar motion by the heading map of
Eq.~\eqref{eq:desired_heading}. By contrast, the slow variable \(S(t)\)
responds on a longer timescale. At \(\lambda_{\rm fb}=10^{-3}\),
\(S(t)\) continues to evolve slowly even after \(M(t)\) and \(Z(t)\) have
nearly saturated. At \(\lambda_{\rm fb}=10^{-2}\), the stronger
perception-to-regulator feedback drives \(S(t)\) more rapidly toward the
high-regulatory branch. This ordering of timescales is the basic
dynamical origin of the memory effects analysed below.

These trajectories also clarify the interpretation of \(S(t)\). The slow
register does not create alignment instantaneously. Instead, it stores a
coarse memory of recent perceptual alignment and subsequently biases the
fast perceptual field through the \(\kappa s_i\) term. Thus, once the
ensemble has approached the ordered branch, the slow variable can sustain
a regulatory bias even when the feedback strength is later reduced.

\subsection{Feedback-induced hysteresis}
\label{sec:hysteresis_results}

To test whether the slow register produces genuine history dependence,
we cyclically sweep the feedback strength
\(\lambda_{\rm fb}\) over the range
\[
10^{-4}\leq \lambda_{\rm fb}\leq 10^{-2}.
\]
The forward branch is obtained by increasing \(\lambda_{\rm fb}\), while
the backward branch starts from the final high-feedback state and then
decreases \(\lambda_{\rm fb}\). At each value of the feedback strength,
the system is first evolved for a relaxation time
\(T_{\rm relax}\), after which the stationary observables are
computed by averaging over a further window \(T_{\rm avg}\):
\begin{equation}
M^\ast(\lambda_{\rm fb})
=
\frac{1}{T_{\rm avg}}
\int_{T_{\rm relax}}^{T_{\rm relax}+T_{\rm avg}}
M(t;\lambda_{\rm fb})\,dt ,
\label{eq:Mstar_numerical}
\end{equation}
and similarly for \(S^\ast\) and \(Z^\ast\).

\begin{figure}[!t]
    \centering
    \includegraphics[width=0.95\linewidth]{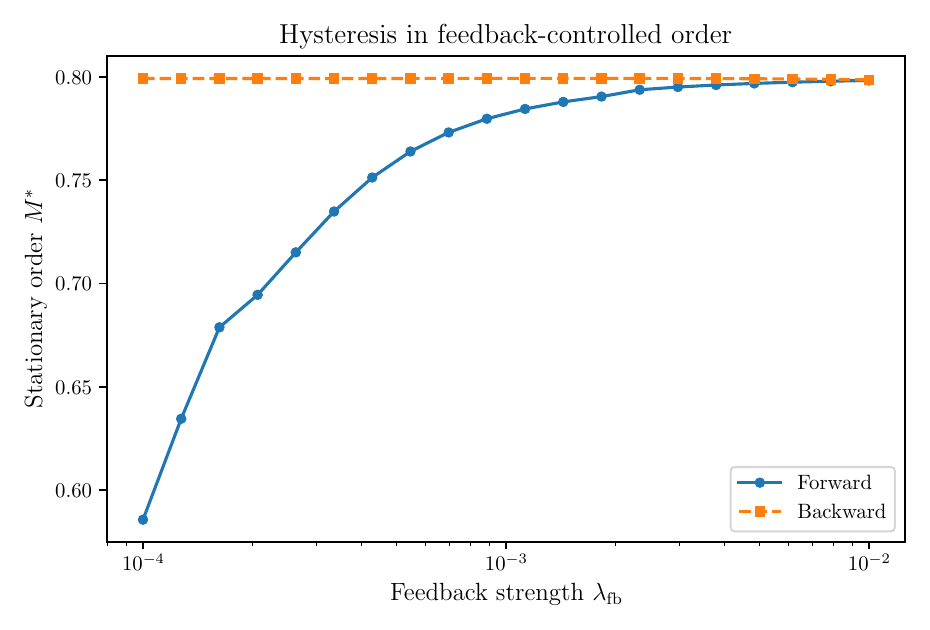}
    \caption{Feedback-induced hysteresis in the polar order parameter.
    The stationary order \(M^\ast\) is plotted during forward and
    backward sweeps of \(\lambda_{\rm fb}\). The backward branch remains
    close to the high-order state even when the feedback is reduced,
    showing that the collective state retains memory of its previous
    regulatory configuration.}
    \label{fig:hysteresis_lambda}
\end{figure}

Figure~\ref{fig:hysteresis_lambda} shows the resulting loop in the
\((\lambda_{\rm fb},M^\ast)\) plane. On the forward branch, the system
starts from a moderately ordered state at low feedback and gradually
approaches the high-order branch as \(\lambda_{\rm fb}\) is increased.
On the backward branch, however, the system remains close to the
high-order state over the entire sweep. Thus the stationary polar order
is not determined by \(\lambda_{\rm fb}\) alone; it also depends on the
history of the slow register. This is the central non-Markovian
signature of the model.

\begin{figure}[!t]
    \centering
    \includegraphics[width=0.88\linewidth]{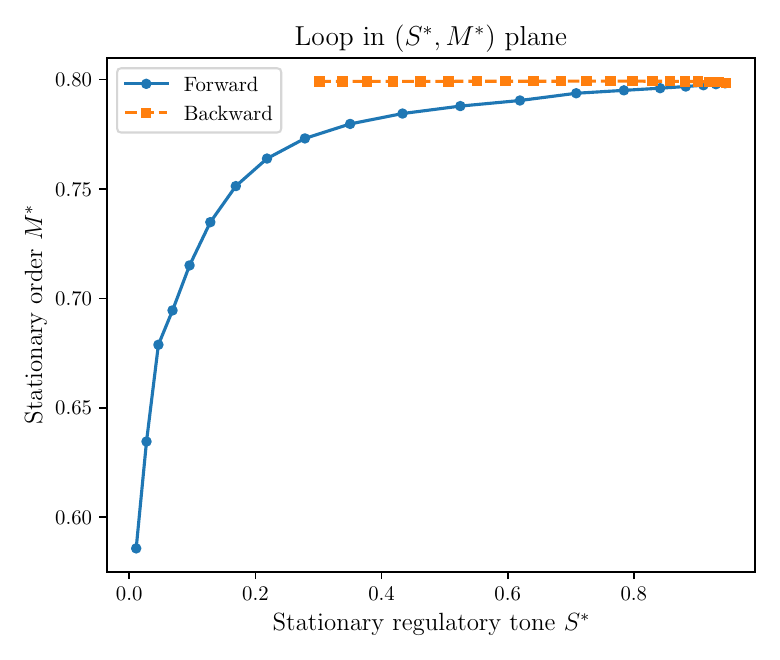}
    \caption{Hysteresis loop in the \((S^\ast,M^\ast)\) plane. The
    forward branch shows gradual co-variation of the regulatory tone and
    polar order. The backward branch remains close to the high-order
    state, indicating persistence of collective alignment after the
    feedback strength is reduced.}
    \label{fig:hysteresis_SM}
\end{figure}

The same hysteresis is displayed in the \((S^\ast,M^\ast)\) plane in
Fig.~\ref{fig:hysteresis_SM}. On the forward branch, \(M^\ast\) rises
together with \(S^\ast\), showing that growth of the slow regulatory tone
amplifies the polar order. On the backward branch, \(M^\ast\) remains
nearly locked to the high-order branch while \(S^\ast\) varies, showing
that the already established collective motion is resistant to reversal.
The loop therefore represents a memory of the prior ordered state stored
in the slow register and expressed at the level of the physical headings.

This behaviour can be understood directly from the slow equation
\eqref{eq:s_dynamics}. Eliminating \(s_i(t)\) from the fast dynamics
would introduce a memory kernel with characteristic time
\(\gamma_s^{-1}\). The forward and backward branches differ because the
hidden slow variable has different values on the two branches even at
the same instantaneous value of \(\lambda_{\rm fb}\). The full system is
Markovian in the enlarged state \((m_x^{ij},m_y^{ij},m_z^{ij},s_i)\),
but the effective dynamics of the physical order parameter \(M(t)\) are
non-Markovian when the slow variable is not observed.

\subsection{Dependence of hysteresis area on the memory timescale}
\label{sec:hysteresis_area_results}

We next quantify how the hysteresis depends on the slow relaxation rate
\(\gamma_s\). For each value of \(\gamma_s\), the same forward/backward
sweep of \(\lambda_{\rm fb}\) is repeated, and the scalar hysteresis
area is computed. 
The scan covers
\[
10^{-4}\leq \gamma_s\leq 10^{-2}.
\]

\begin{figure*}[!ht]
    \centering
    \includegraphics[width=0.5\linewidth]{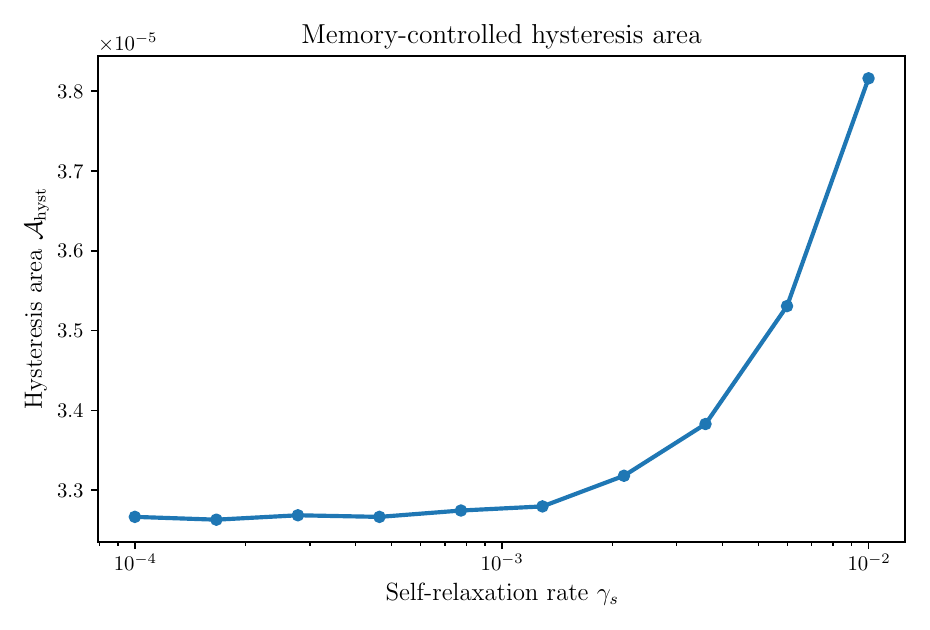}
    \caption{Evolution of hysteresis area with the slow-register
    relaxation rate \(\gamma_s\). The hysteresis area remains finite over
    the scanned range \(10^{-4}\leq\gamma_s\leq10^{-2}\), indicating that
    the collective order retains history dependence across a broad range
    of memory timescales.}
    \label{fig:Ahyst_gamma}
\end{figure*}

Figure~\ref{fig:Ahyst_gamma} shows that the loop area remains finite
throughout the scanned interval. Its magnitude changes only moderately
over most of the range and increases for the largest values of
\(\gamma_s\) used here. The important point for the present work is not
the monotonicity of this finite-window protocol, but the persistence of
a nonzero loop area over two orders of the slow relaxation rate. The
finite area confirms that the forward and backward branches correspond
to distinct internal regulatory histories. In this sense
\(\mathcal A_{\rm hyst}\) is a direct numerical measure of the effective
memory carried by the slow register.

\begin{figure*}[!ht]
    \centering
    \includegraphics[width=0.92\linewidth]{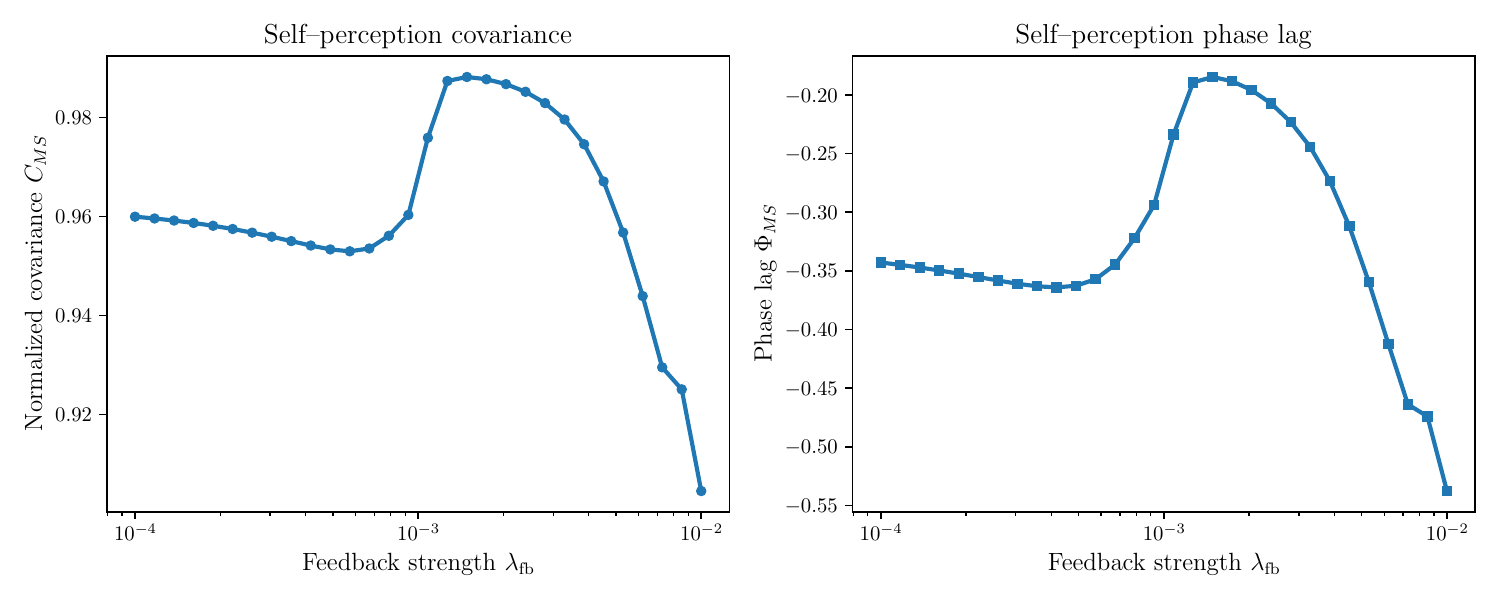}
    \caption{Self--perception coupling as a function of feedback
    strength. Left: normalized covariance \(C_{MS}\) between the polar
    order \(M(t)\) and the slow-register tone \(S(t)\). Right: phase lag
    \(\Phi_{MS}\) obtained from the analytic-signal phase difference.
    Both observables vary non-monotonically with
    \(\lambda_{\rm fb}\), showing that the strongest dynamical
    coordination occurs at intermediate feedback.}
    \label{fig:self_perception_coupling}
\end{figure*}
Because the measurement is performed with finite relaxation and averaging
windows, \(\mathcal A_{\rm hyst}\) should be interpreted as a protocol-dependent
observable: it not only depends on the memory time scale \(\gamma_s^{-1}\), but also on
the sweep schedule, the averaging window, and the coupling parameters.
This is analogous to hysteresis measurements in driven nonequilibrium
systems, where the measured loop area is controlled jointly by the
intrinsic relaxation spectrum and the external protocol. The present
result therefore demonstrates robust memory dependence without requiring
a separate phenomenological memory rule.

\subsection{Self--perception covariance and phase lag}
\label{sec:self_perception_results}

Finally, we examine the dynamical coordination between the physical
order parameter \(M(t)\) and the slow-register tone \(S(t)\). For each
\(\lambda_{\rm fb}\), we run the system for \(T=300\) and compute
\(C_{MS}\) and \(\Phi_{MS}\) over the post-transient half of the
trajectory. The feedback strength is scanned over the same interval
\[
10^{-4}\leq \lambda_{\rm fb}\leq 10^{-2}.
\]

Figure~\ref{fig:self_perception_coupling} shows that both diagnostics
are non-monotonic in \(\lambda_{\rm fb}\). The normalized covariance
\(C_{MS}\) is positive and large throughout the scan, indicating that
the physical order and the regulatory tone are strongly entrained.
However, the entrainment is strongest at intermediate feedback, where
\(C_{MS}\) reaches its maximum. At very weak feedback the slow register
responds only weakly to perceptual alignment. At strong feedback the
system approaches a saturated ordered state, so residual fluctuations
are reduced and the covariance decreases.

The phase lag \(\Phi_{MS}\) shows a corresponding non-monotonic
structure. With the sign convention of Eq.~\eqref{eq:phase_diff}, the
negative values indicate that variations of the slow regulatory tone
lead the corresponding variations in the polar order over the
post-transient window. The lag becomes less negative near the same
intermediate feedback range where \(C_{MS}\) is maximal, indicating
tighter self--perception locking. At larger \(\lambda_{\rm fb}\), the
lag becomes more negative again as the feedback loop approaches a
saturated regime. Thus the self--perception loop is most dynamically
coordinated at intermediate feedback strength, not at either extreme.

The non-monotonicity has a simple interpretation. For small
\(\lambda_{\rm fb}\), perception drives the slow register only weakly,
so \(M(t)\) and \(S(t)\) are correlated but not maximally locked. For
intermediate \(\lambda_{\rm fb}\), the slow register is strongly driven
while remaining dynamically active, producing maximal covariance and the
smallest magnitude of the phase lag. For large \(\lambda_{\rm fb}\), the
system becomes strongly ordered and the slow register approaches
saturation; in that regime, fluctuations are suppressed and the
collective variables have less room to co-vary. This produces the
observed decrease in \(C_{MS}\) and the renewed increase in the magnitude
of \(\Phi_{MS}\).

\subsection{Summary of numerical findings}
\label{sec:results_summary}

The simulations establish four connected results. First, the perceptual
bias \(Z(t)\) and polar order \(M(t)\) rise rapidly, whereas the
slow-register tone \(S(t)\) evolves on a longer timescale. Second,
cyclic variation of \(\lambda_{\rm fb}\) produces distinct forward and
backward branches in \(M^\ast\), demonstrating history-dependent
collective ordering. Third, the hysteresis area remains finite across a
broad scan of \(\gamma_s\), showing that the memory effect is robust to
changes in the slow-register relaxation rate. Fourth, the covariance and
phase lag between \(M(t)\) and \(S(t)\) are non-monotonic in
\(\lambda_{\rm fb}\), revealing an intermediate-feedback regime of
maximal self--perception coordination.

Together these results support the central claim of the paper: a slow
internal regulatory variable coupled to fast perceptual alignment is
sufficient to generate effective non-Markovian collective motion. The
model does not require an explicit delayed alignment rule or an imposed
memory kernel. Instead, memory emerges from the enlarged Markovian
state space through the slow variable \(s_i(t)\), and appears at the
macroscopic level as hysteresis, persistence of the ordered branch, and
nontrivial phase coordination between collective order and regulatory
tone.

\section{Discussion}
\label{sec:discussion}

The numerical results identify slow internal feedback as a distinct
route to memory-dependent collective motion. In the enlarged state space
\((m_x^{ij},m_y^{ij},m_z^{ij},s_i,\hat{\mathbf p}_i,\mathbf x_i)\), the
dynamics are Markovian: the next state is determined by the current
internal variables, headings, and positions. However, when the slow
regulatory variable \(s_i(t)\) is not observed explicitly, the effective
dynamics of the perceptual alignment variables and of the macroscopic
order parameter \(M(t)\) become history dependent. This is made explicit
by the memory-kernel reduction in Appendix~\ref{app:memory_kernel}, and
is observed numerically through the branch separation in
Figs.~\ref{fig:hysteresis_lambda} and \ref{fig:hysteresis_SM}.

The mechanism differs from adding an explicit delay or prescribed memory
kernel to a Vicsek-type update rule. Here the memory variable is an
internal dynamical degree of freedom carried by each agent. It is updated
by recent perceptual alignment and then feeds back into the effective
field \(h_{ij}^{\mathrm{eff}}\) through the slow-to-fast coupling
\(\kappa s_i\). Thus, the hysteresis loop is not imposed at the level of
the macroscopic order parameter. It emerges from a local feedback loop
between fast perceptual alignment and slow regulatory state.

The results also clarify the role of the feedback strength
\(\lambda_{\rm fb}\). Weak feedback is sufficient to produce positive
correlation between the collective order \(M(t)\) and the slow-register
tone \(S(t)\), but the coupling is not maximal in this regime. At
intermediate feedback, the slow register is strongly driven while still
remaining dynamically active, leading to the peak in the normalized
covariance \(C_{MS}\) and the reduced magnitude of the phase lag
\(\Phi_{MS}\). At the largest feedback values studied here, the system
approaches a saturated ordered state. Fluctuations are then suppressed,
and the covariance decreases. This explains why the self--perception
coupling diagnostics in Fig.~\ref{fig:self_perception_coupling} are
non-monotonic rather than simply increasing with feedback strength.

The hysteresis area \(\mathcal A_{\rm hyst}\) should be interpreted as a
protocol-dependent memory observable. It depends not only on the intrinsic
memory time \(\gamma_s^{-1}\), but also on the finite relaxation time,
the averaging window, the feedback sweep range, and the network
topology. The finite values observed in Fig.~\ref{fig:Ahyst_gamma}
therefore demonstrate robust history dependence over the scanned
\(\gamma_s\) range, rather than a universal scaling law. A more complete
finite-rate study would require varying the sweep duration and taking a
quasistatic limit, which we leave for future work.

There are also modelling limitations. First, the simulations use a fixed
topological neighbour graph. This isolates the effect of the internal
slow--fast feedback loop from changes in neighbour identity, but it does
not yet capture all spatial effects of a fully moving active-matter
system with dynamically updated neighbourhoods. Second, the mean-field
closure used to obtain Eq.~\eqref{eq:bloch} neglects connected
correlations between different perceptual channels of the same agent and
between the fast and slow registers. These correlations may become
important near sharp transitions, in sparse networks, or in strongly
heterogeneous populations. Third, the present model uses a single slow
regulatory variable. Biological and engineered collectives may contain
several memory channels operating on different timescales.

Despite these limitations, the model provides a useful minimal
description. It shows that a local internal feedback loop is sufficient
to generate macroscopic memory, hysteresis, and phase-structured
coordination between collective order and internal regulatory tone. This
offers a control principle for engineered swarms, where \(\lambda_{\rm fb}\)
and \(\gamma_s\) could be implemented as programmable gain and memory
parameters, and a phenomenological language for biological collectives,
where analogous slow variables may be inferred from trajectory and
response-time statistics.

\section{Conclusion}
\label{sec:conclusion}

We introduced a two-timescale model of active collective motion in which
each agent carries a fast perceptual register and a slow regulatory register. The fast register determines the instantaneous alignment
weight assigned to neighbouring headings, while the slow one
integrates recent perceptual alignment and feeds back into the fast
alignment field. The internal dynamics were written in a GKSL-derived
Bloch form, used here as an effective positivity-preserving description
of bounded two-state variables rather than as a microscopic quantum
model.

The simulations show that this slow--fast architecture produces
effective non-Markovian collective dynamics. The perceptual bias and
polar order relax rapidly, while the slow regulatory tone evolves on a
longer timescale. Cyclic variation of the feedback strength produces
distinct forward and backward branches in the stationary order parameter,
and the corresponding hysteresis loop persists across a broad range of
slow relaxation rates. The self--perception covariance and phase lag are
non-monotonic in the feedback strength, indicating an intermediate regime
of strongest dynamical coordination between collective order and slow
regulatory tone.

The main conclusion is that memory-dependent collective motion need not
be imposed through an explicit delay or phenomenological memory kernel.
It can emerge from an enlarged Markovian state space when each agent
contains a slow internal variable that stores perceptual history. This
mechanism provides a minimal route from local self-regulation to
macroscopic hysteresis and suggests that internal feedback timescales can
serve as control parameters for collective order in active systems.

\appendix

\section{Bloch-equation derivation from the GKSL generator}
\label{app:bloch_derivation}

This appendix derives the fast-register Bloch equations used in
Eq.~\eqref{eq:bloch}. Consider one perceptual channel \(ij\), and write
its reduced two-state density operator as
\begin{equation}
\rho_{ij}
=
\frac{1}{2}
\left(
I+\mathbf m^{ij}\cdot\boldsymbol{\sigma}
\right),
\qquad
\mathbf m^{ij}
=
(m_x^{ij},m_y^{ij},m_z^{ij}).
\label{eq:app_rho}
\end{equation}
We use the convention that \(m_z^{ij}=+1\) corresponds to the align
alternative and \(m_z^{ij}=-1\) to the not-align alternative.
The effective single-channel Hamiltonian is
\begin{equation}
H_{ij}^{\rm eff}
=
-\Gamma\sigma_x
-
h_{ij}^{\rm eff}\sigma_z ,
\label{eq:app_heff_hamiltonian}
\end{equation}
where \(h_{ij}^{\rm eff}\) is the longitudinal effective field defined in
the main text. The Hamiltonian part of the GKSL equation gives
\begin{equation}
\dot m_\alpha^{ij}
=
{\rm Tr}
\left[
\sigma_\alpha
\left(
-i[H_{ij}^{\rm eff},\rho_{ij}]
\right)
\right],
\qquad
\alpha=x,y,z .
\end{equation}
Equivalently,
\begin{equation}
\dot{\mathbf m}^{ij}\big|_{\rm unitary}
=
2\mathbf B_{ij}\times\mathbf m^{ij},
\qquad
\mathbf B_{ij}
=
(-\Gamma,0,-h_{ij}^{\rm eff}).
\label{eq:app_precession}
\end{equation}
Thus,
\begin{subequations}
\begin{align}
\dot m_x^{ij}\big|_{\rm unitary}
&=
2h_{ij}^{\rm eff}m_y^{ij},
\\
\dot m_y^{ij}\big|_{\rm unitary}
&=
-2h_{ij}^{\rm eff}m_x^{ij}
+
2\Gamma m_z^{ij},
\\
\dot m_z^{ij}\big|_{\rm unitary}
&=
-2\Gamma m_y^{ij}.
\end{align}
\end{subequations}

The dissipative part is represented by one dephasing channel and two
amplitude-relaxation channels,
\begin{equation}
L_\phi
=
\sqrt{\gamma_\phi}\sigma_z,
\qquad
L_-
=
\sqrt{\gamma_-}\sigma_-,
\qquad
L_+
=
\sqrt{\gamma_+}\sigma_+ .
\label{eq:app_lindblad_ops}
\end{equation}
The dephasing channel suppresses the transverse components,
\begin{subequations}
\begin{align}
\dot m_x^{ij}\big|_\phi
&=
-2\gamma_\phi m_x^{ij},
\\
\dot m_y^{ij}\big|_\phi
&=
-2\gamma_\phi m_y^{ij},
\\
\dot m_z^{ij}\big|_\phi
&=
0.
\end{align}
\end{subequations}
The amplitude channels give
\begin{subequations}
\begin{align}
\dot m_x^{ij}\big|_\pm
&=
-\frac{\gamma_-+\gamma_+}{2}m_x^{ij},
\\
\dot m_y^{ij}\big|_\pm
&=
-\frac{\gamma_-+\gamma_+}{2}m_y^{ij},
\\
\dot m_z^{ij}\big|_\pm
&=
-(\gamma_-+\gamma_+)m_z^{ij}
+
(\gamma_+-\gamma_-).
\end{align}
\end{subequations}
Introducing the relaxation times
\begin{equation}
\frac{1}{T_1}
=
\gamma_-+\gamma_+,
\qquad
\frac{1}{T_2}
=
2\gamma_\phi
+
\frac{1}{2T_1},
\label{eq:app_T1_T2}
\end{equation}
and writing the field-dependent longitudinal equilibrium as
\begin{equation}
m_z^{\rm eq}(h)
=
\tanh(\beta h),
\label{eq:app_mzeq}
\end{equation}
the combined unitary and dissipative dynamics become
\begin{subequations}
\label{eq:app_bloch}
\begin{align}
\dot m_x^{ij}
&=
2h_{ij}^{\rm eff}m_y^{ij}
-
\frac{m_x^{ij}}{T_2},
\\
\dot m_y^{ij}
&=
-2h_{ij}^{\rm eff}m_x^{ij}
+
2\Gamma m_z^{ij}
-
\frac{m_y^{ij}}{T_2},
\\
\dot m_z^{ij}
&=
-2\Gamma m_y^{ij}
-
\frac{
m_z^{ij}
-
\tanh(\beta h_{ij}^{\rm eff})
}{T_1}.
\end{align}
\end{subequations}
These are the fast-register Bloch equations used in the main text. The
GKSL construction is used only as a compact positivity-preserving
effective generator for bounded two-state internal variables; no
microscopic quantum dynamics of the active agents is assumed.

\section{Memory-kernel reduction}
\label{app:memory_kernel}

This appendix shows how the slow regulatory variable produces an
effective memory term when it is eliminated from the equations. The slow
variable obeys
\begin{equation}
\dot s_i
=
-\gamma_s(s_i-s_{\rm eq})
+
\lambda_{\rm fb}
(1-s_i^2)
\bar m_z^{\,\mathcal N_i}(t),
\label{eq:app_s_full}
\end{equation}
where
\begin{equation}
\bar m_z^{\,\mathcal N_i}(t)
=
\frac{1}{|\mathcal N_i|}
\sum_{j\in\mathcal N_i}
m_z^{ij}(t)
\end{equation}
is the mean fast-register alignment bias of agent \(i\). The first term
in Eq.~\eqref{eq:app_s_full} relaxes \(s_i\) toward its baseline
\(s_{\rm eq}\) at rate \(\gamma_s\). The second term increases or
decreases \(s_i\) according to the recent alignment state of the fast
register, with strength \(\lambda_{\rm fb}\). The factor
\((1-s_i^2)\) keeps \(s_i\) bounded in \([-1,1]\).

To expose the memory structure, consider the weak-saturation regime
\(|s_i|\ll1\), where \((1-s_i^2)\simeq1\). Then
Eq.~\eqref{eq:app_s_full} becomes the linear equation
\begin{equation}
\dot s_i
=
-\gamma_s(s_i-s_{\rm eq})
+
\lambda_{\rm fb}
\bar m_z^{\,\mathcal N_i}(t).
\label{eq:app_s_linear}
\end{equation}
Solving this equation gives
\begin{equation}
s_i(t)
=
s_{\rm eq}
+
\left[
s_i(0)-s_{\rm eq}
\right]e^{-\gamma_s t}
+
\lambda_{\rm fb}
\int_0^t
e^{-\gamma_s(t-t')}
\bar m_z^{\,\mathcal N_i}(t')\,dt' .
\label{eq:app_s_solution}
\end{equation}
The first two terms describe relaxation toward the baseline. The final
term shows that \(s_i(t)\) stores an exponentially weighted record of
past fast-register alignment. Recent values of
\(\bar m_z^{\,\mathcal N_i}\) contribute strongly, while older values are
suppressed on the timescale \(\gamma_s^{-1}\). Thus
\(\gamma_s^{-1}\) is the memory time of the slow register.

The slow variable enters the fast-register dynamics through the effective
field. For the mean-field form used in the Bloch equations,
\begin{equation}
h_{ij}^{\rm eff}(t)
=
h_{ij}(t)
+
J\bar m_z^{\,\mathcal N_i}(t)
+
\kappa s_i(t),
\end{equation}
substituting Eq.~\eqref{eq:app_s_solution} gives
\begin{align}
h_{ij}^{\rm eff}(t)
&=
h_{ij}(t)
+
J\bar m_z^{\,\mathcal N_i}(t)
+
\kappa s_{\rm eq}
+
\kappa
\left[
s_i(0)-s_{\rm eq}
\right]e^{-\gamma_s t}
\nonumber\\
&\quad
+
\kappa\lambda_{\rm fb}
\int_0^t
e^{-\gamma_s(t-t')}
\bar m_z^{\,\mathcal N_i}(t')\,dt' .
\label{eq:app_heff_memory}
\end{align}
The last term is the memory contribution. It can be written as a
convolution with the causal kernel
\begin{equation}
K(t-t')
=
\kappa\lambda_{\rm fb}
e^{-\gamma_s(t-t')}
\Theta(t-t'),
\label{eq:app_memory_kernel}
\end{equation}
where \(\Theta\) is the Heaviside step function. The effective field at
time \(t\) therefore depends not only on the current fast-register
alignment, but also on its previous values.

This explains the sense in which the dynamics are non-Markovian. If the
full enlarged state
\begin{equation}
\left\{
\mathbf{x}_i,\hat{\mathbf p}_i,s_i,
m_x^{ij},m_y^{ij},m_z^{ij}
\right\}_{i=1,\ldots,N;\,j\in\mathcal N_i}
\end{equation}
is retained, the model is Markovian: the time derivative of every
variable is determined by the current values of these variables. No past
trajectory is needed. However, if \(s_i(t)\) is removed from the state
description, its influence must be replaced by the history integral in
Eq.~\eqref{eq:app_heff_memory}. The reduced dynamics of the fast
perceptual variables alone are therefore effectively non-Markovian.

The memory kernel has amplitude \(\kappa\lambda_{\rm fb}\) and decay time
\(\gamma_s^{-1}\). In the fast-regulator limit, with
\(\lambda_{\rm fb}\) fixed and \(\gamma_s\) large, the kernel becomes
sharply localized and its integrated contribution scales as
\(\lambda_{\rm fb}/\gamma_s\). The slow register then stores little
history, and the reduced dynamics approach an instantaneous-response
limit.

\section{Vicsek-type classical limit}
\label{app:classical_limit}

This appendix shows how the present model reduces to a Vicsek-type
heading update in an appropriate fast-relaxation limit. Consider the
regime of rapid transverse decay and weak slow feedback:
\[
T_2\ll T_1,
\qquad
\kappa s_i\simeq0 .
\]
The transverse components rapidly decay,
\begin{equation}
m_x^{ij}\simeq0,
\qquad
m_y^{ij}\simeq0,
\end{equation}
and the longitudinal equation reduces to
\begin{equation}
\dot m_z^{ij}
=
-
\frac{
m_z^{ij}
-
\tanh(\beta h_{ij}^{\rm eff})
}{T_1}.
\label{eq:app_mz_relax}
\end{equation}
If \(T_1\) is also short compared with the heading-relaxation time
\(\tau_p\), then
\begin{equation}
m_z^{ij}
\simeq
\tanh(\beta h_{ij}^{\rm eff}).
\label{eq:app_mz_adiabatic}
\end{equation}
The heading map
\begin{equation}
\mathbf u_i(t)
=
\rho \hat{\mathbf p}_i(t)
+
\sum_{j\in\mathcal N_i}
w_{ij}
\frac{1+m_z^{ij}(t)}{2}
\hat{\mathbf p}_j(t)
\end{equation}
therefore becomes
\begin{equation}
\mathbf u_i(t)
=
\rho \hat{\mathbf p}_i(t)
+
\sum_{j\in\mathcal N_i}
w_{ij}
\frac{
1+\tanh(\beta h_{ij}^{\rm eff})
}{2}
\hat{\mathbf p}_j(t).
\label{eq:app_weighted_vicsek}
\end{equation}
In the high-sharpness limit \(\beta\to\infty\),
\begin{equation}
\frac{1+\tanh(\beta h)}{2}
\longrightarrow
\Theta(h),
\end{equation}
so each perceptual channel becomes a binary neighbour-following gate.
The desired heading is then a weighted average of the current heading
and selected neighbour headings, followed by normalization through
\[
\hat{\mathbf p}^{\,\rm des}_i
=
\frac{\mathbf u_i}{|\mathbf u_i|+\epsilon}.
\]
This is structurally equivalent to a Vicsek-type local heading update.

The full model goes beyond this classical limit in two ways. First, the
alignment weights are not instantaneous binary gates but dynamical
variables generated by the fast Bloch relaxation. Second, the slow
regulatory state \(s_i(t)\) feeds back into \(h_{ij}^{\rm eff}\), thereby
turning recent perceptual history into a delayed contribution to the
present alignment field. This feedback is the origin of the
memory-dependent ordering and hysteresis discussed in the main text.

\bibliography{ref-pre}
\end{document}